\documentclass[12pt]{article}
\usepackage{graphicx}
\makeatletter
\newcount\@tempcntc
\def\@citex[#1]#2{\if@filesw\immediate\write%
                  \@auxout{\string\citation{#2}}\fi
  \@tempcnta\z@\@tempcntb\m@ne\def\@citea{}\@cite{\@for\@citeb:=#2\do
    {\@ifundefined
       {b@\@citeb}{\@citeo\@tempcntb\m@ne\@citea%
                   \def\@citea{,}{\bf ?}\@warning
       {Citation `\@citeb' on page \thepage \space undefined}}%
    {\setbox\z@\hbox{\global\@tempcntc0\csname b@\@citeb%
                     \endcsname\relax}%
     \ifnum\@tempcntc=\z@ \@citeo\@tempcntb\m@ne
       \@citea\def\@citea{,}\hbox{\csname b@\@citeb\endcsname}%
     \else
      \advance\@tempcntb\@ne
      \ifnum\@tempcntb=\@tempcntc
      \else\advance\@tempcntb\m@ne\@citeo
      \@tempcnta\@tempcntc\@tempcntb\@tempcntc\fi\fi}}\@citeo}{#1}}
\def\@citeo{\ifnum\@tempcnta>\@tempcntb\else\@citea\def\@citea{,}%
  \ifnum\@tempcnta=\@tempcntb\the\@tempcnta\else
   {\advance\@tempcnta\@ne\ifnum\@tempcnta=\@tempcntb%
     \else \def\@citea{--}\fi
    \advance\@tempcnta\m@ne\the\@tempcnta\@citea\the\@tempcntb}\fi\fi}
\makeatother

\begin{document}

\begin{center}
{\bf Dynamic and Quasistatic Trajectories in Quasifission Reactions
and Particle Emission }

\smallskip\ 

V. P. Aleshin$^1$, M. Centelles$^2$, X. Vi\~nas$^2$, and N. G.
Nicolis$^3$ 

$^1$Institute for Nuclear Research, Kiev, 252028, Ukraine

$^2$Departament d'Estructura i Constituents de la Mat\`eria, Facultat
de \\
F\'{\i}sica, Universitat de Barcelona, Diagonal 647, E-08028
Barcelona, Spain

$^3$ Department of Physics, University of Ioannina, Ioannina 45110,
Greece 

\bigskip\ 

Abstract
\end{center}

We show that the quasifission paths predicted by the one-body
dissipation dynamics, in the slowest phase of a binary reaction,
follow a quasistatic path, which represents a sequence of states of
thermal equilibrium at a fixed value of the deformation coordinate.
This establishes the use of the statistical particle-evaporation model
in the case of dynamical time-evolving systems. Pre- and post-scission
multiplicities of neutrons and total multiplicities of protons and
$\alpha $ particles in fission reactions of $^{63}$Cu+$^{92}$Mo,
$^{60}$Ni+$^{100}$Mo, $^{63}$Cu+$^{100}$Mo at 10 MeV/u and
$^{20}$Ne+$^{144,148,154}$Sm at 20 MeV/u are reproduced reasonably
well with statistical model calculations performed along dynamic
trajectories whose slow stage (from the most compact configuration up
to the point where the neck starts to develop) lasts some $35\times
10^{-21}$ s.



%

\section{Introduction}

In the last years many experimental efforts have been devoted to the
study of heavy-ion fission at beam energies below 10--20 MeV/u
accompanied with the emission of light particles. Initial experiments
involved the observation of fission fragments emitted in coincidence
with neutrons \cite{Gavron,Hinde89,Hinde92} or charged-particles \cite
{Lindl,Lacey,Benrachi,Boger,P1,P2}. More recently, combined
coincidence neutron and charged-particle data became available
\cite{Gonin,L,Lou,T}.  The most important result of these experiments
consists in the estimation of fission times which appear to be
$10^3$--$10^4$ times longer than the characteristic time of the
single-particle motion. This fact implies that thermal equilibrium is
being established over the intrinsic degrees of freedom at each value
of the fission coordinate $Q$ and validates the use of the Rayleigh
dissipative  function technique to treat the coupling of $Q$ with the
particle  degrees of freedom \cite{Landau}.

Heavy-ion fission reactions accompanied by particle emission are used
as a testing ground of the dissipation mechanisms of large-scale
collective motion in hot nuclei (for a review see Refs.\
\cite{AARS,Froebrich}). The main reason is that this process is
simpler than other types of deep-inelastic collisions. The simplicity
originates from the fact that fission is not sensitive to the fusion
stage of a binary reaction whose understanding is far from clear.

When mass and projectile energy increase, the possibility to isolate
the ground-state to scission-point motion becomes problematic because
of the increasing contribution of quasifission, a fission-like process
occurring at angular momenta $J$ exceeding some value $J_0$ where the
fission barrier disappears. Quasifission in a clear cut sense has been
observed in very heavy systems \cite{Toke.85} which do not have a
fission barrier even at $J=0$ $\hbar$. These experiments provided a
motivation for the creation of the code HICOL which proved to be
capable of reproducing the general features of the process
\cite{Feld}. 

The Rayleigh function used in HICOL is built within the one-body
dissipation model, proposed in Ref.\ \cite{Blocki}. It combines the
wall formula in the mononuclear regime with an improved version of the
so-called `completed window formula' \cite{RS} in the dinuclear stage.
Calculations show that quasifission originates from the fact that the
potential of the composite system for $J\approx $ $J_0$ is very flat.
Combined with strong friction this gives the system enough time to
thermalize the relative velocity of the colliding nuclei and to relax
the mass-asymmetry mode.

In medium mass reaction systems, a quasifission configuration can
execute several rotations before scission, so that it is impossible to
disentangle quasifission from fusion-fission by using fragment angular
distributions as in heavier systems. Moreover, due to decreasing in
angular momentum caused by the emission of   few particles, a pocket
can emerge in the initially flat potential, and a quasifission
trajectory will go over into a fusion-fission one.

The introduction of quasifission trajectories to fission-evaporation
routines is not an easy task. It requires answers on some principal
questions. For example, in the fission case it is assumed
\cite{Gontchar} that, for all $J$ within the fusion-fission angular
momentum window, the system moves along the bottom of the fission
valley of a non-rotating nucleus. Thus a question arises on whether the
quasifission trajectories go along this path. Another related question
is whether the evaporation model is applicable for particle emission
along the quasifission trajectories.

The main purpose of the present work is to elucidate these questions.
As a result we shall get a better perspective for a uniform
description of fusion-fission and quasifission reactions accompanied
by particle emission. To illustrate this opportunity, we test the
compatibility of the charged particle clock with the neutron clock in
calculations which effectively account for fusion-fission and
quasifission trajectories. The study is performed in the $A\approx $
160, $E_x$= 200--300 MeV region, where a quite complete set of data on
multiplicities of fission-associated light particles is available
\cite{Gonin,Lou}. 

The paper is organized as follows. In Sect.~2 we define the
quasistatic path and in Sect.~3 we introduce the geometrical
quantities needed to characterize the quasistatic and dynamic
configurations of the system. These quantities facilitate the
comparison between the dynamic and quasistatic paths in Sect.~4. In
Sect.~5 we perform Monte Carlo simulations of light-particle
evaporation along dynamic trajectories and compare the results with
available experimental data. Our conclusions are drawn in the last
section. 

%

\section{The Quasistatic Path}

The fundamental quantity in the construction of fission models is the
total energy $\widetilde{E}\{\rho \}$ of the nucleus expressed as a
functional of the spatial nucleon density $\rho $. The definition of
this functional and its general properties are well documented (e.g.,
see  Refs.\ \cite{Pet91,PP,Br85,Cen90}). Given $\widetilde{E}\{\rho \}$ 
one can use the equations 
\begin{equation}
\label{dEdr} \frac{ \delta \widetilde{E} \{\rho\} }{\delta \rho} =0,
\qquad \int \rho ({\bf r}) d{\bf r}=A,
\end{equation}
where $\delta/\delta \rho $ is the functional derivative, to find the
ground state and the saddle point densities $\rho _{{\rm gs}}$ and
$\rho _{ {\rm sd}}$. The difference between the corresponding
energies,  $\widetilde{E} \{\rho _{{\rm sd}}\}-\widetilde{E}\{\rho
_{{\rm  gs}}\}$, gives the fission barrier height.

Since nuclei are leptodermous objects the terms `density' and `shape'
are often used synonymously. As indicated first in Ref.\ \cite{S62},
the so-called conditional equilibrium shapes may play an important
role in nuclear dynamics. To define these shapes, one introduces a
quantity $Q$ characterizing the elongation of the nucleus. $Q$ is a
functional of the density and has the form \begin{equation}
\label{constr}Q\{\rho \}=\int q({\bf r})\rho ({\bf r})d{\bf r},
\end{equation}  where $q({\bf r})$ is a known function. The density
$\rho _{Q^{*}}$ of the conditional equilibrium shape is the solution
of equations (\ref{dEdr}) in the class of densities $\rho $ restricted by
the  condition $Q\{\rho \}=$ $Q^{*}$ where $Q^{*}$ is a constant.
As pointed out in Ref.\ \cite{S62}, the sequence of conditional
equilibrium shapes may acquire a physical meaning, for instance it can
describe the process of fission if $Q$ is changing adiabatically
compared to all other degrees of freedom. It was conjectured in Ref.\
\cite{S62} that the adiabaticity condition is likely to be satisfied
in heavy systems where $\widetilde{E}\{\rho \}$ changes from the
ground state to the saddle point by a few MeV only.

Nowadays, there is a strong evidence \cite{Froebrich} that the
dynamical equation for fission of hot nuclei should take into account
the coupling of $ Q$ with the single-particle degrees of freedom,
which  causes a slowing down of the fission time scale to values
exceeding  the single-particle times by a few orders of magnitude.
This allows one to assume that at each instant of time the
single-particle degrees of freedom are in thermal equilibrium under
the constraint of the known $Q$ and the respective velocity $\dot Q$.
Expressed in formal terms, this means that the entropy $S$ of the
system is maximal for the given constraints:
\begin{equation}
\label{dsdr} \frac{\delta S}{\delta \rho} =0, \qquad
\int \rho  ({\bf r})d{\bf r }=A, \qquad Q=Q^{*}, \qquad 
 \dot Q=\dot  Q^{*}.
\end{equation}
Equations (\ref{dsdr}) define a state of partial thermal
equilibrium \cite{Landau}, or quasistatic state, for short.

Denoting the intrinsic excitation energy by $E_x$ and using the Fermi
gas formula for the entropy
\begin{equation}
S=2\sqrt{a\,(E_x-\widetilde{E}\{\rho \}+\widetilde{E}\{\rho_{\rm gs}\})}\,, 
\end{equation} 
one can find the density $\rho $ of the quasistatic state by looking
for the conditional minimum of $\widetilde{E}$. 
This is justified by
the fact that the variation of the level density parameter $a$ with
shape is very smooth in comparison to that of $\widetilde{E}$
\cite{Garcias}. By changing $Q^{*}$ one gets a sequence of quasistatic
states which we call the quasistatic path. The fact that fission
proceeds along the quasistatic path leads to significant
simplifications in the formal description of this process. It also
allows us to employ the statistical-model treatment of particle
emission for the fissioning nucleus, because the single-particle
degrees of freedom are in thermal equilibrium at each point of the
quasistatic path.

Another process in which the quasistatic shapes can be useful is
quasifission. The complete thermal equilibrium of the single-particle
motion at a given shape may be reached during the reseparation stage
of a quasifission reaction if in the fusion stage the mass asymmetry
mode and the relative velocity of the two colliding nuclei have
relaxed. As a result, the quasifission trajectories in the
reseparation stage will get on the quasistatic path. In practical
terms, quasifission reactions are described by the one-body
dissipation model of heavy ion collisions \cite{Feld} implemented in
the code HICOL, while the conditional equilibrium densities can be
calculated in the framework of the extended Thomas--Fermi (ETF) model
of non-spherical nuclei \cite{Garcias,Garc,Cent}. With these two
models we wish to verify whether the dynamic trajectories in
quasifission reactions follow the quasistatic path.

Our ETF calculations will be confined to mass and axially symmetric
(about the $z$ axis) prolate density distributions $\rho (r,z)$
normalized to the mass number $A$, where $r=\sqrt{x^2+y^2}$ and $x$,
$y$, $z$ are the Cartesian coordinates. The energy density of the ETF
model incorporates second-order gradient corrections with spin-orbit
and effective mass terms \cite{Br85}, which are very important in
describing the nuclear surface. We have performed the ETF calculations
using a realistic Skyrme interaction, namely SkM$^{*}$ \cite{Br85}.
From the ETF--SkM$^{*}$ functional, we obtain fully self-consistent
nuclear densities by solving the associated variational
Euler--Lagrange equations in cylindrical coordinates, imposing a given
value of the quadrupole moment $Q_2 $ \cite{Garcias,Garc,Cent}:
\begin{equation}
Q_2=2\pi \int_0^\infty \!\!\int_{-\infty }^\infty 
\left[ 2z^2-r^2\right] \,\rho (r,z)\,r\,drdz\,.
\end{equation}
To account for nuclear rotation with angular momentum $J$ we
have included a rotational energy
\begin{equation}
E_{{\rm rot}}=\frac{J^2}{2I}
\end{equation}
to the ETF energy functional. Here, $I$ is the rigid-body moment 
of inertia 
\begin{equation}
I=\pi \int_0^\infty \!\!\int_{-\infty }^\infty \left[ 2z^2+r^2\right]
\,m\,\rho (r,z)\,r\,drdz\,,
\end{equation}
where $m$ is the nucleon mass. It was assumed that the spin axis is
directed perpendicular to the symmetry axis of the compound nucleus.

In HICOL the constant density approximation is used. Therefore, the
nuclear density is completely determined by the profile function
$y(z)$ whose rotation about the symmetry axis $z$ generates the
nuclear surface. This poses difficulties in the comparison between
dynamic and quasistatic densities, which can nevertheless be avoided
by  introducing some generalized characteristics of the nuclear
densities. 

%

\section{Geometry of the Composite System}

For the sake of comparison between dynamic and quasistatic paths we
introduce the elongation coordinate $D_{{\rm mm}}$ and the neck
coordinate $R_{{\rm neck}}$. In axially symmetric nuclei these
quantities are given by
\begin{equation}
\label{Dmm}D_{{\rm mm}}={\frac{8\pi }A}\int_0^\infty \!\!\int_0^\infty
\!z\,\rho (r,z)\,r\,drdz 
\end{equation}
and 
\begin{equation}
\label{Rneck}R_{{\rm neck}}^2={\frac 2{\rho _0}}\int_0^\infty r\,\rho
(r,z=0)\,dr\,. 
\end{equation}

The elongation coordinate $D_{{\rm mm}}$ defines the distance between
the centers of mass of the two halves of the nucleus. It was used by
Strutinsky \cite{S62} as a constraint operator in the
integro-differential equation for the profile function $y(z)$ of
leptodermous nuclei. Our definition (\ref{Rneck}) of the neck radius
is obtained from the requirement that a nucleus with constant density
$\rho _0$ and a geometrical neck radius equal to $R_{{\rm neck}}$, has
the same number of particles in the cross section of its neck as the
nucleus having the distributed density $\rho $. In the following,
$\rho _0$ in Eq.\ (\ref{Rneck}) will be identified with the one used
in the code HICOL, namely $\rho _0=A/(\frac 43\pi R_0^3)$ where
$R_0=1.18\,A^{1/3}$ fm. From Eqs.\ (\ref{Dmm}) and (\ref{Rneck}) one
finds that for a spherical nucleus with a constant density $\rho _0$,
the elongation $D_{{\rm mm}}$ is equal to $\frac 34R_0$ and that the
neck radius is $R_{{\rm neck}}=R_0$.

In HICOL the profile function $y(z)$ of the composite system is
parameterized by two spheres smoothly connected by a second-order
curve \cite {Blocki.80}. For these so-called Blocki shapes we define
$D_{{\rm mm}}$ as the distance between the centers of mass of the two
parts of the nucleus on both sides of a plane $z=z_{{\rm m}}$. For
$z_{{\rm m}}$ we take the mean value of the left and right matching
points ($z_{{\rm m}}=0$ for symmetric shapes). We identify $R_{{\rm
neck}}$ of the Blocki profile with $y(z_{{\rm m}})$.

For mass symmetric shapes, the Blocki profile function reads 
\begin{equation}
\label{y}y^2(z)=\left\{ 
\begin{array}{ll}
R_1^2-(z+{s/2})^2 & \mbox{for\ }-R_1-{s/2}\le z\le -z_1\,, 
\\ \alpha +\beta
z^2 & \mbox{for\ }-z_1\le z\le z_1\,, \\ R_1^2-(z-{s/2})^2 
& \mbox{for\ } z_1\le z\le {s/2}+R_1\,. 
\end{array} \right. 
\end{equation}
Given the volume $V_0$ of the nucleus, the parameters $R_1$, $z_1$,
$\alpha $ and $\beta $ can be expressed in terms of the two collective
degrees of freedom $s$ and $\sigma $, where $s$ is the distance
between the centers of the spheres and
\begin{equation}
\label{sigma}\sigma =\frac{V_0-8\pi R_1^3/3}{V_0} 
\end{equation}
is a measure of the constriction of the system. 

Indeed, from Eq.\ (\ref{sigma}) one obtains 
\begin{equation}
\label{R1}R_1=R_0\left( {\frac{1-\sigma }2}\right) ^{1/3},
\end{equation}
where $R_0$ is the radius of a spherical nucleus of volume $V_0$. If
we  equate the values of $y(z)$ and its derivative on both sides of
$z=z_1$ and require that the total volume of the shape generated by
$y(z)$ equals $V_0$, we get
\begin{eqnarray}
\alpha & = & R_1^2+\frac s2\left( z_1-\frac s2\right) ,
\label{alpha} 
\\[2mm]
\beta &=& {s\over 2z_1}-1 \,,
\label{beta}
\\[2mm]
z_1 &=& \frac s2\sqrt{1-G} \,,
\label{z1}
\end{eqnarray}
with 
\begin{equation}
\label{Gs0}G=\frac 3{s_0^2}-\frac 2{s_0^3}
\,{\frac{1+\sigma }{1-\sigma }}\,, \qquad s_0=\frac s{2R_1}\,.
\end{equation}

For the Blocki shapes defined by Eq.\ (\ref{y}), Eq.\ (\ref{Dmm}) yields 
\begin{equation}
\label{DmmBS}D_{{\rm mm}}=\frac 3{4R_0^3}\,\left\{ \frac{s^4}{48}\left[
(1-G)^{3/2}-1\right] +{\frac 12}s^2R_1^2+{\frac 43}sR_1^3+R_1^4\right\} . 
\end{equation}
The value of $R_{{\rm neck}}^2$ coincides with $y^2(0)$ and according to
(\ref{y}) is equal to $\alpha $. Thus, using Eqs.\ (\ref{alpha}) and 
(\ref{z1}) we obtain 
\begin{equation}
\label{RneckBS}R_{{\rm neck}}^2=R_1^2+\frac{s^2}4\,\left( \sqrt{1-G}
-1\right) . 
\end{equation}
Given $R_0$, Eqs.\ (\ref{DmmBS}) and (\ref{RneckBS}) together with
Eqs.\ (\ref{R1}) and (\ref{Gs0}) allow one to express $D_{{\rm mm}}$
and $R_{{\rm neck}}$ in terms of $s$ and $\sigma $. 

In the following section, instead of the quadrupole moment, we use sometimes
the moment of inertia as a constraint operator.
It should be noted that for mass symmetric Blocki shapes  
these quantities can be found analytically:
\begin{equation}
\label{Q2an}Q_2=4\pi \rho _0\left[ \frac{h(R_1+h)^3}4\left( R_1-\frac
h3\right) -\frac h6(R_1^2-h^2)z_1^2-{\frac 2{15}}h^2z_1^3+\frac{hz_1^4}{20}
\right] , 
\end{equation}
\begin{eqnarray}
I=\pi m\rho _0\left[ \frac{(R_1+h)^3}5\left( {\frac
83}R_1^2- 
\frac{R_1h}2+\frac{h^2}6\right)  
\right.
\nonumber\\ 
\left.
 +\frac h3(R_1^2-h^2)z_1^2+{\frac 4{15}}
h^2z_1^3+\frac{hz_1^4}{30}\right] ,
\label{Ian}
\end{eqnarray}
where $h=s/2$.

%

\section{Dynamic and Quasistatic Paths}

We now describe calculations performed in order to compare the dynamic
trajectories with the quasistatic path. The calculations are carried
out in the $A\approx 160$ composite mass region, a region of
continuous experimental efforts
\cite{Gavron,Hinde89,Lindl,Lacey,Boger,P1,P2,Gonin,Lou}. We start with
the quasifission reaction following a $^{60}$Ni$+^{100}$Mo collision
at the beam energy $E=600$ MeV\@. In Fig.\ 1 we display the
equidensity contour plots corresponding to the sequence of the ETF
densities of conditional equilibrium for $^{160}$Yb, which represents
the composite system in the collision. For each ETF density shown we
have calculated the values of $D_{{\rm mm}}$ and $R_{{\rm neck}}$ by
means of Eqs.\ (\ref{Dmm}) and (\ref {Rneck}). 
Inserting these $D_{{\rm mm}}$ and $R_{{\rm neck}}$ into Eqs.\ (\ref{DmmBS}) 
and (\ref{RneckBS}) and solving these equations with respect to $s$ and 
$\sigma $, we can prescribe the Blocki profiles to the density distributions. 
From Fig.\ 1 one can see that such profiles are close, in general, to the 
${\frac 12}\rho _0$ curves of the ETF density distributions. 
In dinuclear configurations, however, the nascent fragments predicted by 
the ETF model are somewhat flattened in the $r$ direction compared to the 
spherical form.

The densities shown in Fig.\ 1 were calculated accounting for the
rotational energy with angular momentum $J$= 86 $\hbar $. The reasons
for this choice of $J$ will become clear in the next section. The
calculations indicate that the quasistatic path depends weakly on the
angular momentum. When one goes from $J$=86 $\hbar $ to $J$=0 $\hbar $
the quasistatic $D_{{\rm mm}}$ and $ R_{{\rm neck}}$ change at most by
0.1 fm. In earlier ETF calculations \cite{Garcias} the energy
$\widetilde{E}$ of the rotating nucleus along the quasistatic path was
found to be very close to the sum of the energy of the non-rotating
nucleus and the rotational energy computed with the moment of inertia
of the latter. Moreover it was shown that the level density parameter
is affected by rotation in a negligible amount if $Q_2$ is fixed.
These findings were interpreted as an indication that rotation has a
small influence on the nuclear densities calculated in conditional
equilibrium. Our direct calculations of shape parameters along the
quasistatic path agree with this conclusion.

We repeated the calculation of the $J$= 86 $\hbar $ quasistatic path
with the constraint on the moment of inertia $I$, instead of $Q_2$.
The path in the $(D_{{\rm mm}},R_{{\rm neck}})$ space turned out to be
almost the same as the one we had found with the $Q_2$ constraint: the
differences in $R_{{\rm neck}}$ are no larger than $\sim $1\%.
Imposing the $I$ constraint we not only obtain the same values for
$Q_2$ as with the $Q_2$ constraint, but also find that the
hexadecapole moment $Q_4$ along the path is very similar. This means
that the nuclear shapes must be equivalent with the $I$ or $Q_2$
constraint, as well.

The dynamic evolution of the shape of the $^{60}$Ni$+^{100}$Mo
composite system (at $J$=86 $\hbar $ and $E$= 600 MeV) computed with
HICOL is shown in Fig.\ 2. In units of $10^{-21}\,{\rm s}$, the first
stage (when neck fills in) takes about 0.2, the mononucleus lives
about 30 and the scission stage lasts about 5. At the end of the first
stage the individual temperatures and angular velocities of the two
nuclei, predicted by HICOL, become approximately equal. The time
dependence of $D_{{\rm mm}}$ for different $J$ values is depicted in
Fig.\ 3, which shows a clear separation of the reactions into the
fusion stage taking a fraction of $10^{-21}\,{\rm s}$ and a much
longer reseparation stage. The duration time of the latter strongly
changes from one $J$ to another. For example, this time decreases from
$45\times 10^{-21}\,{\rm s}$ to $15\times 10^{-21}\,{\rm s}$ when $J$
increases from 85 $\hbar$ to 100 $\hbar$.

Figure 4 shows on a ($D_{{\rm mm}},R_{{\rm neck}}$) plot how the
dynamic trajectories with $J$=70, 75, 80, 85, 90, 95, 100, 105 $\hbar
$ are joining the quasistatic path determined for $J$= 86 $\hbar $.
The systems with $J$ =70 and 75 $\hbar $ terminate at the different
points of this path (in the case of $J$=75 $\hbar $ this happens after
a slight rebound) while those with $J$ =80--105 $\hbar $, having got
on the quasistatic path shortly after rebound, proceed towards the
scission point. In the vicinity of the scission point, they start to
deviate progressively from the quasistatic path and yield a thinner
neck for the same value of the elongation. We have found the dynamic
paths, in the reseparation phase, to be rather stable against
variations in the excitation energy and mass asymmetry. This follows
from our dynamic calculations performed for the system
$^{60}$Ni+$^{100}$Mo at the beam energy of 1200 MeV (Fig.\ 5), and
for $^{48}$Ca on $^{112}$Sn at the beam energy of 480 MeV (Fig.\ 6).

Very small but regular deviations of the dynamic trajectories in their
slowest phase from the quasistatic path are clearly observed in Figs.\
4, 5 and 6. They are probably related to the fact that the dynamical
and variational calculations involve different forces, and that in the
ETF calculation we varied the whole density rather than just the
nuclear profile. To verify this assumption we calculated the
quasistatic path for $J$= 86 $ \hbar $, with a $Q_2$ constraint, using
the Yukawa-plus-exponential (YPE) forces \cite{Krappe} which are
employed in the code HICOL\@. We looked for the conditional minimum of
the system energy in the class of mass symmetric Blocki shapes. The
result is displayed on Fig.\ 6 with square symbols. According to this
figure, the dynamical paths in their slowest part practically coincide
with the quasistatic YPE path.

It is interesting to note that the YPE quasistatic path predicts the
onset of the scission stage at noticeably smaller values of $D_{{\rm
mm}}$ than the Skyrme path. This is consistent with the fact that the
saddle point configuration for the YPE force is more compact than for
the SkM$^{*}$ force. The deformation energy for the YPE force reaches
its maximum (of about 27.8 MeV) at $D_{{\rm mm}}\sim 13.6$ fm, whereas
the SkM$^{*}$ energy reaches its maximum (about 25.1 MeV) at $D_{{\rm
mm}}\sim 15.6$ fm. These deviations manifest the scale of errors
introduced into variational calculations by simple parameterizations
of nuclear shapes.

%

\section{Particle Emission}

In the preceding section we have shown that the dynamic trajectories
in  the reseparation stage closely follow the quasistatic path. This
means  that statistical models of particle emission can be applied in
this  stage. Below, we describe the technique for light-particle
evaporation  calculations along the slow phase, which will allow us to
perform  comparisons with experimental data.

Our procedure is based on a Monte Carlo simulation of particle decay
chains in nuclei with a time-dependent shape. Given the excitation
energy $E_x$, angular momentum $J$, the dynamic path and the time
$t_{\rm i}$ for the beginning of the slow phase, we calculate the
neutron ($R_{\rm n}$), proton ($R_{\rm p}$) and alpha particle
($R_\alpha $) emission rates. We assume that the emission times
$t_{\rm e}$ are distributed according to the exponential law $\exp
(-R_{\rm tot}t_{\rm e})$, where $R_{\rm tot}= R_{\rm n}+R_{\rm
p}+R_\alpha $ is the total emission rate. A specific value of $t_{\rm
e}$ is chosen using a generator of exponentially distributed random
numbers.  Having sampled the type of the emitted particle in
proportion to the weights $R_{\rm n}/R_{\rm tot}$, $R_{\rm p}/R_{\rm
tot}$ and $R_\alpha /R_{\rm tot}$, we find the average excitation
energy $\bar u$ and root mean square angular momentum $j_{{\rm rms}}=
\sqrt{\overline{j^2}}$ of the corresponding daughter nucleus and take
the latter as the new decaying nucleus. Assuming its shape to be the
same as that of the parent nucleus at $t_{\rm i}+t_{\rm e}$, we
simulate the next decay. The chain of decays is terminated when the
emission time exceeds the scission time of the nucleus and one
proceeds to the next chain.

The pre-scission multiplicities $M_\nu (J)$ are calculated as
\begin{equation} \label{MnuJ}M_\nu (J)={\frac 1{{\cal N}}}
\sum_{i=1}^{{\cal N}}{\cal N}_\nu (i), \end{equation} where ${\cal
N}_\nu (i)$ is the number of particles of type $\nu $ ($\nu =$ n, p,
$\alpha$) in a decay chain classified with the index $i$. Here, ${\cal
N }$ is the total number of decay chains for a given $E_x$ and $J$.
The explicit expressions for $R_\nu $, $\bar u$, $\overline{j^2}$ are
summarized in the Appendix. They were obtained in the framework of the
classical statistical model of particle emission from non-spherical
nuclei \cite{Aleshin88,Aleshin90,Aleshin93}. As shown in Ref.\
\cite{Aleshin96}, the effects of the shape distortions on particle
emission are treated by these formulas more rigorously than in
heuristic models \cite {Blann,Dossing,Aj,Nicolis,Pomorski}.

The input parameters of these formulas are the effective separation
energies $S_\nu ^{{\rm eff}}$ calculated including deformation
energies \cite{L}, the level density parameter $a$, the height
$V_{{\rm b}}$ and the radius $R_{ {\rm b}}$ of the corresponding
spherical barrier experienced by a particle. In the following
calculations we employ the YPE values of $S_\nu ^{{\rm eff} } $ for
$\nu =$ n, p, $\alpha$. In Fig.\ 7 the YPE values (solid lines) are
compared with the ETF separation energies (dashed lines) on the
quasistatic path in $^{160}$Yb. The ratio $I/I_0$ of the deformed
nucleus moment of inertia to that of the spherical one is used as the
coordinate along the path.

The level density parameters $a$ to be used later have been normalized
at the spherical shape to the experimental value $A/8.8$ MeV$^{-1}$
obtained in Ref.\ \cite{Henss}. The shape dependence of $a$ has been
calculated with the YPE forces following the prescription of T\~oke
and Swiatecki \cite{TS}. As seen from Fig.\ 8, these $a$ are close, by
magnitude and shape dependence, to the $a$ values from the ETF method.
The dependence of $V_{{\rm b}}$ and $R_{{\rm b}}$ for n, p, $\alpha $
on $A$ and $Z$ is parameterized in the same way as in Ref.\
\cite{Blann1}. The energy of the emitted neutron entering the
corresponding $R_{{\rm b}}$ is replaced with its mean value $\approx
2\tau $, where $\tau $ is the temperature of the daughter nucleus.

Recently, Lou {\it et al} \cite{Lou} measured multiplicities of light
particles in fission reactions of 10 MeV/u $^{63}$Cu+$^{92,100}$Mo and
20 MeV/u $^{20}$Ne+$^{144,148,154}$Sm. Our analysis of these data
along with the data of Gonin {\it et al} \cite{Gonin} on
$^{60}$Ni+$^{100}$Mo is presented in Fig.\ 9 and Table 1. The
measurements of multiplicities of n, p, $\alpha $ in the latter work
were performed at 9.2 and 10.9 MeV/u. The data we analyze are obtained
by interpolation of the reported values to 10 MeV/u.

In the analysis we took into account that in these reaction systems an
appreciable amount of particles escape from the system during the {\em
pre-equilibrium} stage. The measurements of linear-momentum transfer
from projectile to target allowed Lou {\it et al} \cite{Lou} to
estimate that the mass and charge removed in this stage are $(\delta
A,\delta Z)$= (8,4), (8,4), (6,3), (7,3) and (6,3) in the reactions
$^{63}$Cu+$^{92,100}$Mo and $^{20}$Ne+$^{144,148,154}$Sm, respectively.
Since experimental information on $\delta A$ and $\delta Z$ for the
$^{60}$Ni+$^{100}$Mo system is not available, we used the values
$(\delta A,\delta Z)$=(8,4) in the closest system,
$^{63}$Cu+$^{100}$Mo. Lou {\it et al} \cite{Lou} estimated the initial
excitation energies of the equilibrated compound nuclei to be $E_x=$
227, 267, 289, 277 and 282 MeV in the reactions of 10 MeV/u
$^{63}$Cu+$^{92,100}$Mo and 20 MeV/u $^{20}$Ne+$^{144,148,154}$Sm,
respectively. For the system of 10 MeV/u $^{60}$Ni+$^{100}$Mo, the
linear interpolation between the excitation energies $E_x$= 251 MeV
and 293 MeV at 9.2 MeV/u and 10.9 MeV/u, respectively, obtained in
Ref.\ \cite{Gonin}, results in $E_x$= 271 MeV\@.

This information on $\delta A$, $\delta Z$ and $E_x$ was used as input
in our {\em pre-scission} (equilibrium) multiplicity calculations. For
all systems we used $t_{\rm i}= 1.1\times 10^{-21}$s and simulated
${\cal N}= 200$ chains of decays. For each system we took only one
HICOL trajectory, namely the trajectory whose contact-to-scission time
$t_{{\rm cs}}$ is closest to $ 40\times 10^{-21}$ s. This condition
results in $J$= 78, 86, 86, 85, 88, 93 $ \hbar $ in the reactions of
$^{63}$Cu+$^{92}$Mo, $^{60}$Ni+$^{100}$Mo, $^{63} $Cu+$^{100}$Mo at 10
MeV/u and $^{20}$Ne+$^{144,148,154}$Sm at 20 MeV/u, respectively. The
slow stage of the so-chosen (`effective') trajectories lasts some
$35\times 10^{-21}$ s. This is consistent with the fission time scale
of $(35\pm 15)\times 10^{-21}$ s deduced from a systematic study of
pre-scission neutron multiplicities in 27 fission reactions induced by
$^{16,18}$O, $^{40} $Ar and $^{64}$Ni on targets with $A=141$--238
\cite{Hinde92}. 

In the {\em post-scission} emission calculations, the thermal energy
of the composite system at the moment of scission was shared between
the complementary fragments in proportion to their masses. The spins
of the fragments (about 6 $\hbar$ per fragment) were taken from HICOL
output. To find $A$ and $Z$ of the primary fragments, we used the
calculated n, p and $\alpha$ pre-scission multiplicities (see Table
1).  The calculated post-scission multiplicities of neutrons in all
reactions but $^{63}$Cu+$^{92}$Mo are confined between the value of
3.7$\pm $0.4 measured in the system of $^{16}$O+$^{154}$Sm at $E_x$=
206 MeV \cite{Hinde92} and the value of 4$\pm $1.1 for
$^{60}$Ni+$^{100}$Mo at 10 MeV/u which follows from interpolation of
the 9.2 and 10.9 MeV/u data ($3.6\pm1$ and $4.5\pm1.2$, respectively)
of Ref.\ \cite{Gonin}.

With accounting for post-scission emission, which is essential, in
fact, only for neutrons, the total (equilibrium) multiplicities appear
to be within the likely systematic uncertainties of the experimental
points. The only noticeable exceptions are the total proton
multiplicities in $^{63}$Cu+$^{92} $Mo and $^{60}$Ni+$^{100}$Mo, when
the measured values are smaller than the calculated ones by factors of
1.6 and 2.1, respectively. With the exception of these two data
points, the overall agreement of the calculations with the rest of the
data signifies a consistency of the neutron with the charged particle
clock.  The low proton multiplicities in $^{60}$Ni+$^{100}$Mo
reactions  observed in Ref.\ \cite{Gonin} have motivated further
experimental  studies. In a recent work, Charity {\it et al}
\cite{Charity} studied the nearby system $^{64}$Ni+ $^{100}$Mo at a
similar excitation energy and found 2--3 times greater multiplicities
of {\em fusion}-associated p and $\alpha$ compared to those of Ref.\
\cite{Gonin}. 

It should be noted that no fitting parameters were used in our
analysis. The employment of the transmission coefficients for p and
$\alpha$ from Ref.\ \cite{MA} would destroy the quality of the
description by strongly enhancing the $\alpha $ emission. For
instance, the pre-scission multiplicities in the $^{60}$Ni+$^{100}$Mo
system become 6.73, 2.29 and 1.81 for n, p and $\alpha$,
respectively, instead of 7.27, 2.11 and 0.98 obtained with the
transmission coefficients from Ref.\ \cite{Blann1}.

%

\section{Discussion and Conclusion}

To shed more light on the role of quasifission trajectories in the
reactions of our study, it is useful to estimate the fusion-fission
$J$-window and compare it with available experimental information on
the angular momenta, associated with the evaporation residue and
fission cross sections. Nuclei emerging at the end of the evaporation
cascades undergo fission if their angular momentum is confined between
the angular momentum where the fission barrier height $B_{\rm f}(J)$
equals the neutron separation energy and the angular momentum where
$B_{\rm f}(J)$ vanishes \cite{CPS,AWOA}. To reconstruct the
corresponding $J$-window in the beginning of the evaporation cascades,
one has to account for the angular momentum removed by pre-scission
particles. 

In the case of $^{60}$Ni+$^{100}$Mo at 10 MeV/u, we find that the
fusion-fission $J$-window at the end of the evaporation cascades is
57--82 $\hbar$, on the average. Light particles evaporated along the
effective trajectory, remove on the average about 21 $\hbar $.
Therefore, we estimate the corresponding $J$-window in the beginning
of the {\em equilibrium} emission stage, i.e.\ for $^{152}$Dy, to be
78--103 $\hbar $. Since $B_{\rm f}(J)$ for $^{152}$Dy vanishes at 83
$\hbar$, trajectories within this window, with the exception of those
with $J$=78--83 $\hbar $, belong to quasifission.

It is interesting to note that the width of the so-defined $J$-window
is close to the `total' (fusion-fission plus quasifission) $J$-window
widths in the closest systems to ours, where data are available. A
fission $J$-window of 70--103 $\hbar $ and 49--67 $\hbar $ is implied
in studies of $^{40}$Ar+$^{109}$Ag at 8.4 MeV/u \cite{Britt} and
$^{20}$Ne+$^{159}$Tb at 16 MeV/u \cite{Keutgen}, respectively. Thus we
may conclude that the majority of fission-evaporation reactions start
on quasifission trajectories which after losing angular momentum end
up on trajectories in the potential with a non-zero fission barrier.

To summarize, the present work was motivated by the desire to
estimate  the perspective for the inclusion of quasifission into
fission-evaporation  codes. Towards this aim we tested whether the
quasifission trajectories, in the reseparation stage, follow the
fission path. Our calculations were performed in the $A\approx $ 160
region at the bombarding energy of 10--20 MeV/u.

For the sake of comparison between different shapes (assumed to be
axially symmetric) we use the two-dimensional space of the collective
variables describing elongation and constriction. The elongation is
characterized by the distance between the two halves of the nucleus.
The constriction is described by the neck radius defined in a way
applicable for distributed densities. In this collective space, we
first check the sensitivity of the quasistatic path to the angular
momentum of the system and to the form of the constraint operator. The
quasistatic paths found for different angular momenta (including those
exceeding the critical angular momentum for fission) are practically
indistinguishable from the one with $J$= 0 $\hbar $. The moment of
inertia as a constraint operator was found to generate a sequence of
shapes which coincides with the one obtained using the constraint on
the quadrupole moment.

The same space of collective variables was used to compare the
sequences of shapes predicted by the HICOL code at different values of
the entrance-channel angular momentum. The parts of the quasifission
trajectories corresponding to the slow stages of the evolution show
the eventual convergence to the quasifission trajectory with the
minimal $J$. This latter is found to follow closely the quasistatic
path obtained with ETF model. This close coincidence of dynamic
trajectories with the quasistatic path was found to occur in a wide
range of bombarding energies and for quite different entrance-channel 
mass asymmetries.

Since the dynamics of quasifission reactions is well described by
quasistatic paths, at least during the slowest phase, the statistical
evaporation model can be applied for the description of particle
emission from such systems. This follows from the fact that at each
point of the quasistatic path the system reaches thermal equilibrium.

Using single {\it effective} dynamical trajectories whose slow stage
lasts some $35\times 10^{-21}$ s, it was made possible to reproduce
reasonably well experimental data on pre- and post-scission
multiplicities of neutrons and total multiplicities of protons and
$\alpha $-particles emitted from thermally equilibrated systems. This
agreement was achieved without any {\it ad hoc} statistical model
parameter adjustments and shows a consistency between the neutron and
charged particle clock. The duration time of the slow stage of the
employed dynamical trajectories was found consistent with the
results of systematic studies.

%

\section*{Acknowledgments}

This research was supported by Grants No.\ PB98-1247 from the DGICYT
(Spain), 1998SGR-00011 from the DGR (Catalonia), the General
Secretariate of Research and Technology of the Ministry of Industry,
Research and Technology of Greece ($\Pi ENE\Delta $ '95 No.\ 696), and
a NATO Fellowship Program for the Year 1996-97. Useful comments of
Profs.\ W. Swiatecki and P. Schuck are greatly appreciated. We are
thankful to Dr.\ Th. Keutgen for communicating us information on 
experimental data before publication.


\section*{Appendix}

In the following we outline the formalism we used for the calculation
of decay rates for statistical particle emission from equilibrated
compound nuclei. The excitation energy, angular momentum, deformation
energy and moment of inertia of the parent nucleus are denoted as
$E_x$, $J$, $E_{{\rm def}}$ and $I_x$, respectively. Given these
quantities, the thermal energy of the parent nucleus is defined by
$$
q_x=E_x-\frac{J^2}{2I_x}-E_{{\rm def}}. 
$$
Its reduced level density is 
$$
\omega _x(q_x)=\frac 1{t_x^4(I_xa_x)^{\frac 32}}\exp \left[
2\sqrt{a_xq_x} \right] , 
$$
where 
$$
t_x=\frac 3{4a_x}+\sqrt{\left( \frac 3{4a_x}\right) ^2+\frac{q_x}{a_x}}, 
$$
and $a_x$ is the level density parameter. Similar formulas are used
for the reduced level density $\omega (q)$ of the daughter nucleus
with thermal energy $q$. Its moment of inertia and level density
parameter are denoted by $I$ and $a$.

The mass, spin and effective separation energy of the emitted particle
are denoted as $m_\nu $, $s_\nu $ and $S_\nu ^{{\rm eff}}$,
respectively. Let $z_{{\rm matter}}$ be half a length of matter
distribution in the deformed shape, and let $R_{\rm b}$ and $R_{{\rm
matter}}$ be the barrier radius and matter radius for the spherical
shape. Then half a length of the figure generated by the barrier line
is postulated to be
$$
z_0=z_{{\rm matter}}+(R_{\rm b}-R_{{\rm matter}}).
$$
Given the matter profile function $y=y(z)$ and the Coulomb potential
$\Phi(z)$ along this profile, the barrier line $\rho (z)$ and the
potential barrier $U(z)$ along this line were calculated from
$$
\rho (z)=Ky\left( \frac zK\right) ,
\qquad U(z)=\frac{V_{\rm b}}{\Phi _0}\Phi
\left( \frac zK\right) , 
$$
where $K=z_0/z_{{\rm matter}}$ is the scaling factor, $V_{\rm b}$ is
the $s$-wave potential barrier in the spherical nucleus and $\Phi _0$
is the Coulomb potential on the edge of the spherical matter
distribution. 

The key characteristics of the residual nucleus are its average
thermal energy $\bar q$ and temperature $\tau $ corresponding to this
thermal energy. These quantities are calculated from
$$
\bar q\approx E_x-S_\nu ^{{\rm eff}} - E_{{\rm def}} 
- \frac{J^2}{2I}-V_{\rm b}, $$
$$
\tau =\frac 2a+\sqrt{\left( \frac 2a\right) ^2+\frac{\bar q}a}. 
$$

With these definitions the particle emission rate is given by 
$$
R_\nu =\frac{2s_\nu +1}{2\pi }m_\nu z_0^2\tau ^2\frac{\omega (\bar q)}
{\omega _x(q_x)}\exp \left[ \frac 1\tau \left( E_x-S_\nu ^{{\rm
eff}}-E_{{\rm def}}-\frac{J^2}{2I}-\bar q\right) \right] G(\tau ,b), 
$$
where 
$$
G(\tau ,b)= \int_{-1}^1d\zeta \sqrt{\eta ^2+\eta ^2 
\left( \frac{d\eta} {d\zeta}\right) ^2}
\exp \left[ -\frac{U(\zeta z_0)}\tau
+ b \zeta ^2+\frac 12 b \eta ^2 \right]
I_0 \left(\frac 12b\eta ^2\right),
$$
$$
b=b(\tau )=\frac{m_\nu z_0^2}{2\tau }\left( \frac JI\right) ^2, 
\qquad \eta =\eta (\zeta )=\frac{\rho (\zeta z_0)}{z_0}, 
$$
and $I_0(x)$ is a Bessel function of the first kind of imaginary
argument.

The averaged square of the angular momentum of the daughter nucleus is
$$
\overline{j^2}=J^2-4I\tau b\frac{d\ln G(\tau ,b)}{db}. 
$$
Given this quantity and $\bar q$, the average excitation energy of the
daughter nucleus reads 
$$
\bar u=\bar q+\frac{\overline{j^2}}{2I}+u_{{\rm def}} , 
$$
where $u_{{\rm def}}$ is the daughter nucleus deformation energy.

The above expressions give approximate values of $R_\nu$, $\overline{j^2}$ and
$\bar u$ because they use approximate values of $\bar q$ and $\tau $. 
To correct these latter quantities we find a new value of $\bar q$ using the 
formula  
$$
\bar q=E_x-S_\nu ^{{\rm eff}}-E_{{\rm def}}-\frac{J^2}{2I}-2\tau -\tau
^2 \frac{d\ln G(\tau ,b(\tau ))}{d\tau } 
$$
at the initial value of $\tau$ and insert it into the formula for $\tau$. 
This procedure was iterated until convergence of $\tau $ was attained.

%

\newpage

Table 1. Comparison between experimental and calculated light-particle
multiplicities for the reaction systems studied in Refs.\ \cite{Gonin} and 
\cite{Lou}. The neutron number $N$ of the composite system after the
pre-equilibrium emission stage is indicated. From left to right the systems
are 10 MeV/u $^{63}$Cu+$^{92}$Mo, $^{60}$Ni+$^{100}$Mo, $^{63}$Cu+$^{100}$Mo
and 20 MeV/u $^{20}$Ne+$^{144,148,154}$Sm

\begin{center}
\begin{tabular}{lcccccc}
\hline
$N$ & 80 & 86 & 88 & 89 & 92 & 99 \\
\hline
n$^{\rm pre}$ & 3.75 & 7.27 & 7.46 & 7.65 & 8.2 & 10.87 \\ 
n$^{\rm post}$ & 1.79 & 3.84 & 3.87 & 3.82 & 3.89 & 3.97 \\ 
n$^{\rm tot}$ & 5.54 & 11.11 & 11.33 & 11.47 & 12.09 & 14.84 \\ 
n$^{\rm tot,exp}$ & 6(1) & 11.25(2.7) & 10.9(1.5) & 10.3(1.3) & 
12.2(1.6) & 16.9(1.2) \\ 
p$^{\rm pre}$ & 2.97 & 2.11 & 1.97 & 2.65 & 1.85 & 0.82 \\ 
p$^{\rm post}$ & 0.18 & 0.06 & 0.05 & 0.07 & 0.05 & 0.02 \\ 
p$^{\rm tot}$ & 3.15 & 2.17 & 2.02 & 2.72 & 1.9 & 0.84 \\
p$^{\rm tot,exp}$ & 2.02(0.3) & 1.03(0.15) & 1.6(0.3) & 2.14(0.3) &
1.83(0.5) & 0.97(0.5) \\ 
$\alpha^{\rm pre}$ & 1.17 & 0.98 & 0.93 & 1.24 & 1. & 0.66 \\
$\alpha^{\rm post}$ & 0.03 & 0.02 & 0.01 & 0.02 & 0.01 & 0.01 \\
$\alpha^{\rm tot}$ & 1.2 & 1 & 0.94 & 1.26 & 1.01 & 0.66 \\
$\alpha^{\rm tot,exp}$ & 1(0.3) & 0.77(0.11) & 1.69(0.4) & 1.3(0.3)
& 0.7(0.4) & 0.63(0.4) \\
\hline
\end{tabular}
\end{center}

\newpage

%

\section*{Figure Captions} \mbox{}

Fig.\ 1. Equidensity contours for $^{160}$Yb calculated by the ETF
method at the indicated values of the quadrupole moment, with angular
momentum $J$= 86 $\hbar $. From outside to inside the lines represent
contours of constant density $\rho =$ 0.1, 0.3, 0.5, 0.7, 0.9 and 1.1,
in units of $\rho _0$. The dashed curves represent the Blocki profile
with the parameters $s$ and $\sigma $ calculated in terms of $D_{{\rm
mm}}$ and $R_{{\rm neck}}$ obtained from the corresponding ETF density
distribution. 

Fig.\ 2. Time evolution of the $^{60}$Ni$+^{100}$Mo system from the
code  HICOL at $E=600$ MeV and $J$= 86 $\hbar $. The time is indicated
in units of $10^{-21}\,{\rm s}$.

Fig.\ 3. Time dependence of $D_{{\rm mm}}$ for the reaction
$^{60}$Ni+$^{100}$Mo at $E= 600$ MeV and $J$= 79-105 $\hbar$
predicted by the HICOL code.

Fig.\ 4. Dynamic (for $J= 70$, 75, 80, 85, 90, 95, 100, 105 $\hbar$)
and quasistatic (for $J$= 86 $\hbar$) paths in the
$^{60}$Ni+$^{100}$Mo collision at 600 MeV are shown as solid and
dashed lines, respectively. The spherical shape is shown as a cross.
Numbers along the curves indicate the time in units of $10^{-21} \,
{\rm s}$ for the dynamic path at $J$=86 $\hbar$. The insert shows 
the final stage of the fusion trajectories with $J$=70 $\hbar$ and 
$J$=75 $\hbar$.

Fig.\ 5. Dynamic paths (for $J=70$, 75, 80, 85, 90, 95, 100, 105
$\hbar$) in the $^{60}$Ni$+^{100}$Mo collision at 1200 MeV are shown
as solid lines. The quasistatic path for $J$= 86 $\hbar $ is
represented by the dashed line. The spherical shape is shown as a
cross. 

Fig.\ 6. Dynamic paths (for $J$= 70, 75, 80, 85, 90, 95, 100, 105
$\hbar $) for the collision $^{48}$Ca$+^{112}$Sn at 480 MeV are shown
as solid lines. The dashed line represents the quasistatic path
corresponding to the self-consistent ETF variational calculation with
the SkM$^{*}$ force. The closed square symbols show the quasistatic
path found for the YPE forces in the space of Blocki shapes.

Fig.\ 7. The effective separation energies of neutrons, protons and
alpha particles along the quasistatic path in $^{160}$Yb. The solid
lines show the calculation for the dynamic shapes with YPE forces. The
dashed lines indicate the ETF effective separation energies.

Fig.\ 8. The level density parameter along the quasistatic path. The
solid line represents the $a$ values normalized to $A/8.8$ MeV$^{-1}$
for the spherical shape with the T\~oke-Swiatecki shape-dependent factor
based on YPE forces. The dashed line indicates the ETF calculation.

Fig.\ 9. Equilibrium total multiplicities of neutrons, protons and
alpha particles (pre-scission plus post-scission) as a
function of the neutron number {\em N} of the emitting system (after
the pre-equilibrium stage). Experimental data points from Refs.\
\cite{Lou} and \cite{Gonin} are shown by filled squares and crosses,
respectively. The short-dashed lines represent the calculated
pre-scission (equilibrium) multiplicities. The solid lines show the
total calculated multiplicities. In the top (neutron) panel, the open
circles represent the total measured multiplicities accounting for
pre-equilibrium emission \cite{Lou}. From left to right the systems are
10 MeV/u $^{63}$Cu+$^{92}$Mo, $^{60}$Ni+$^{100}$Mo,
$^{63}$Cu+$^{100}$Mo and 20 MeV/u $^{20}$Ne+$^{144,148,154}$Sm.

%
\newpage

\begin{center}
\includegraphics{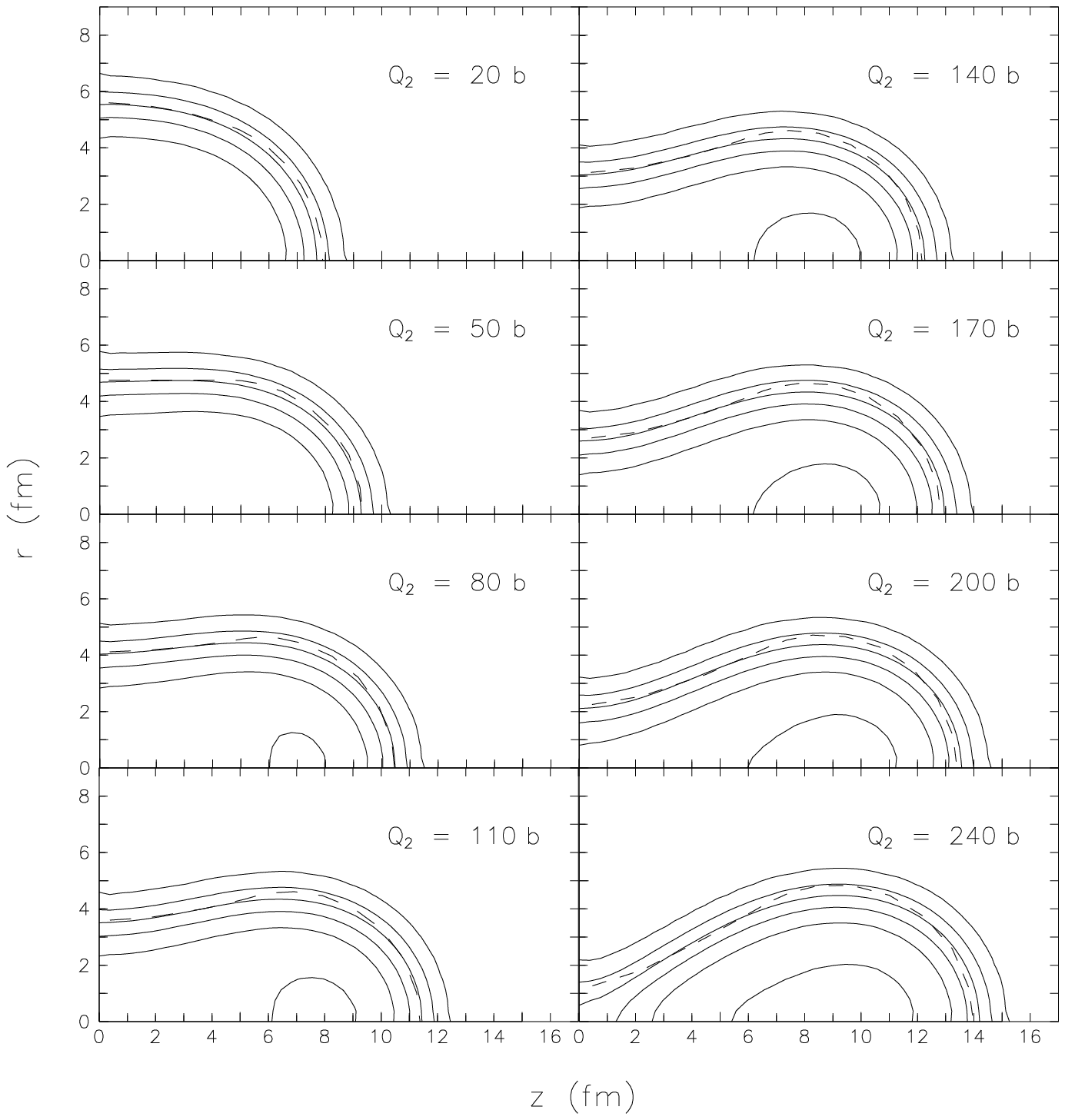}

\vspace*{2cm} Figure 1
\end{center}


\begin{center}
\includegraphics{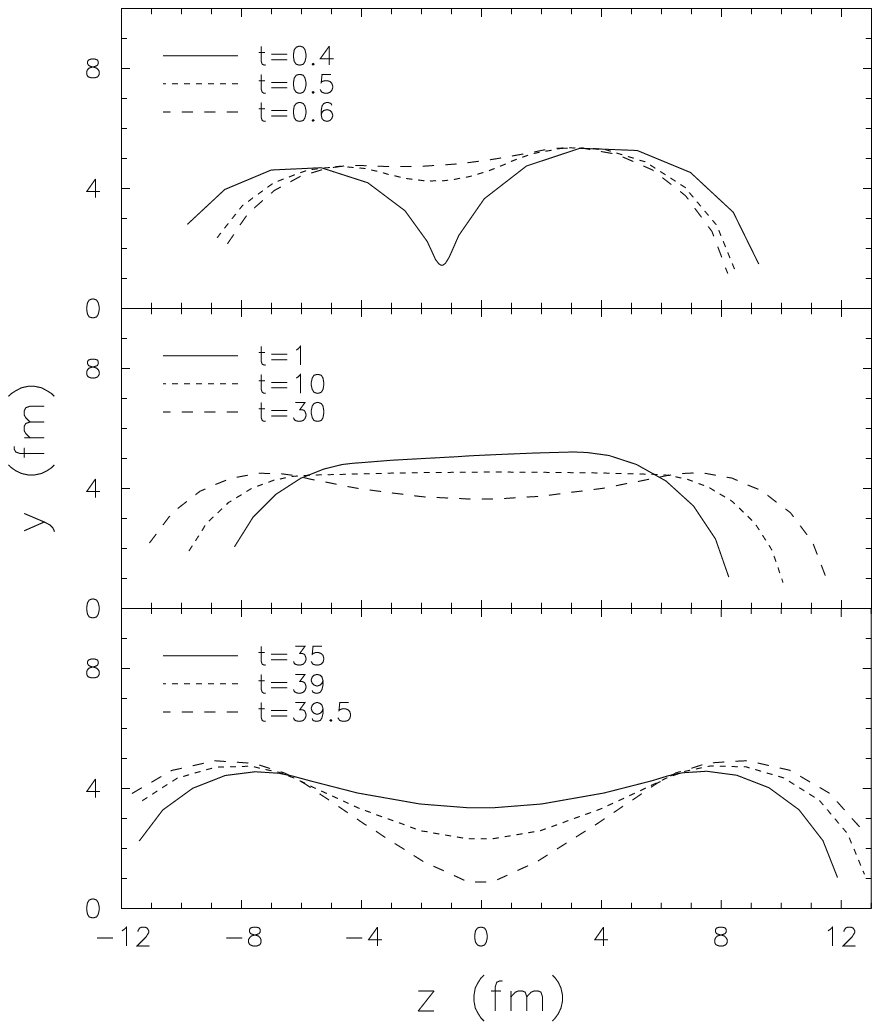}

\vspace*{2cm} Figure 2
\end{center}

\begin{center}
\includegraphics{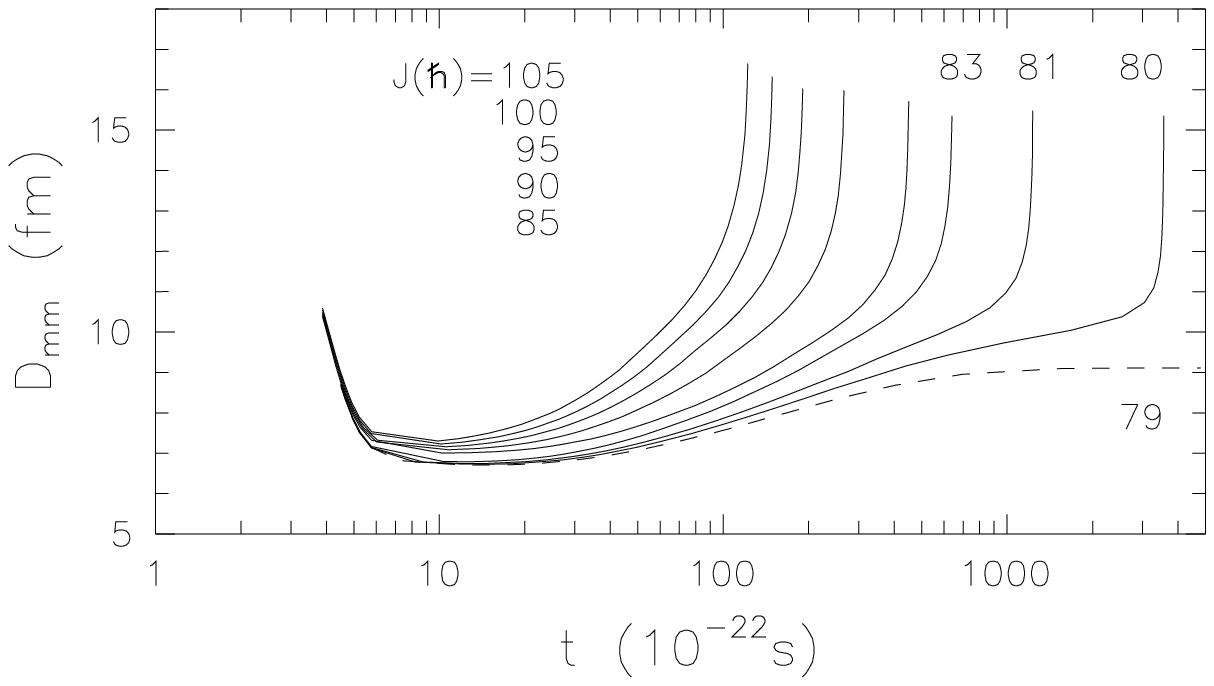}

\vspace*{1cm} Figure 3
\end{center}

\vspace*{1cm}

\begin{center}
\includegraphics{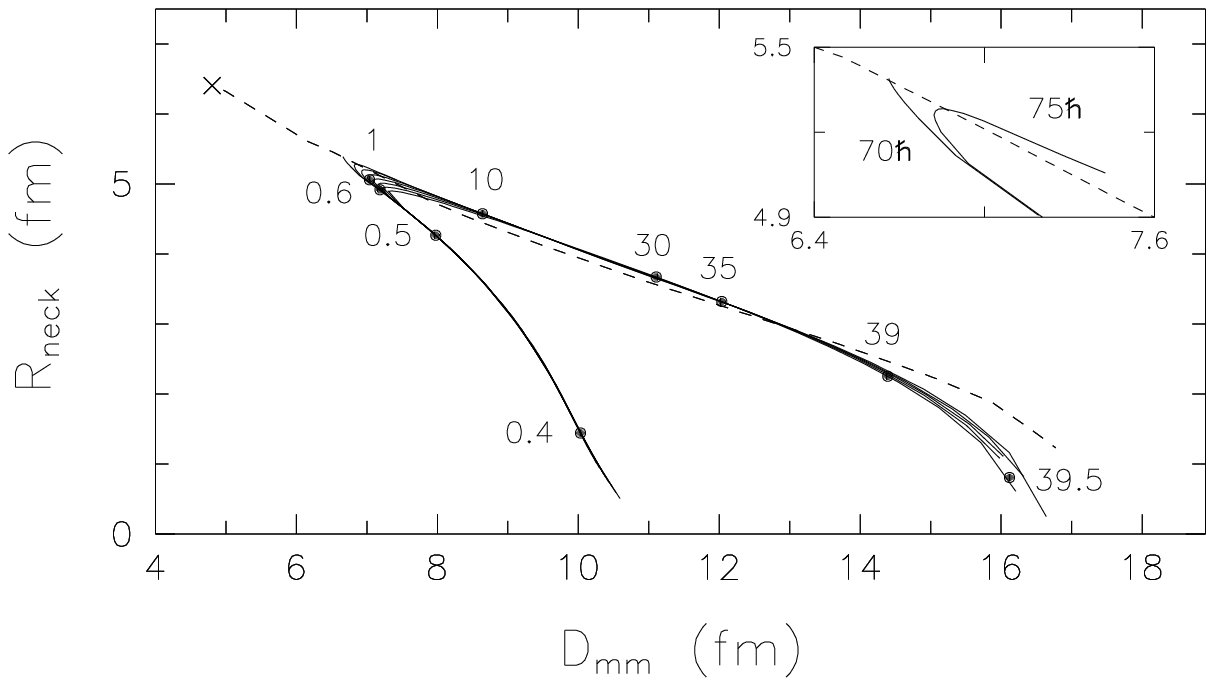}

\vspace*{1cm} Figure 4
\end{center}

\begin{center}
\includegraphics{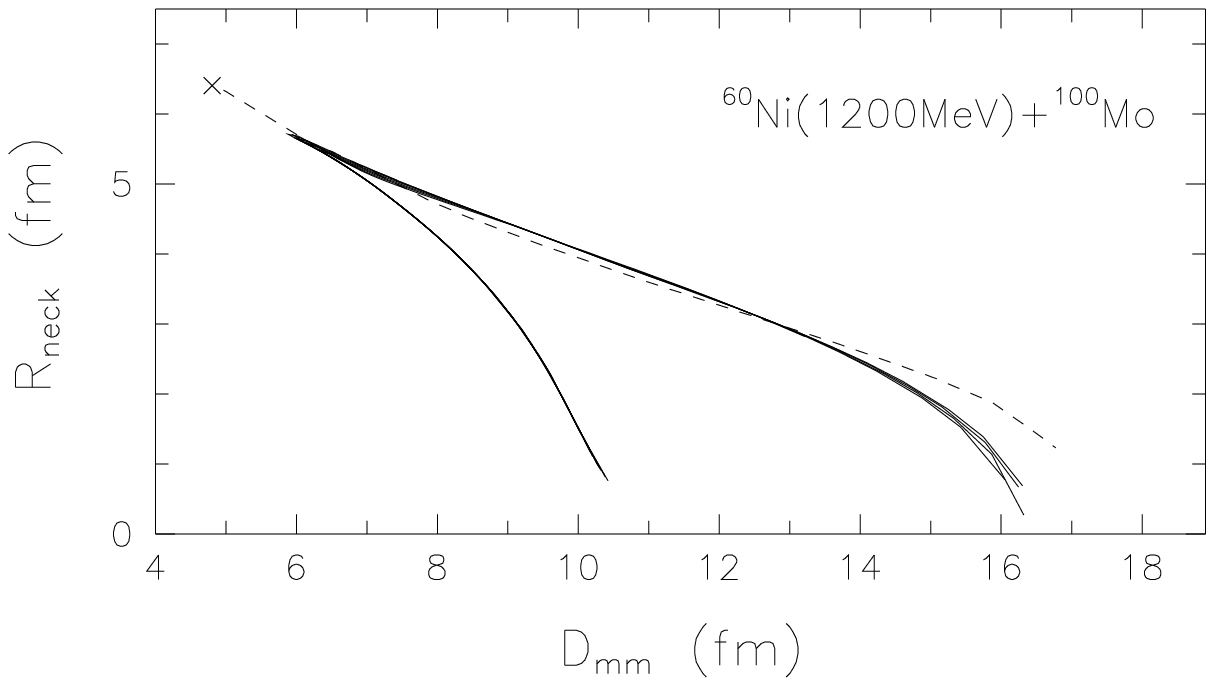}

\vspace*{1cm} Figure 5
\end{center}

\vspace*{1cm}

\begin{center}
\includegraphics{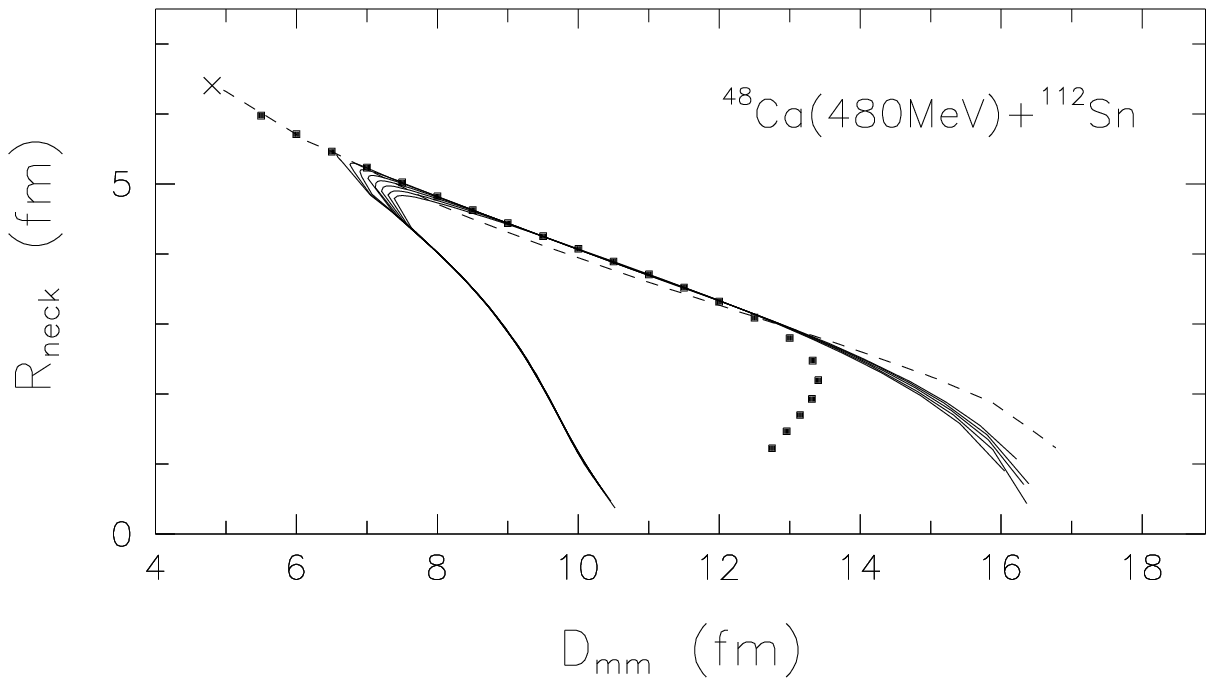}

\vspace*{1cm} Figure 6
\end{center}

\begin{center}
\includegraphics{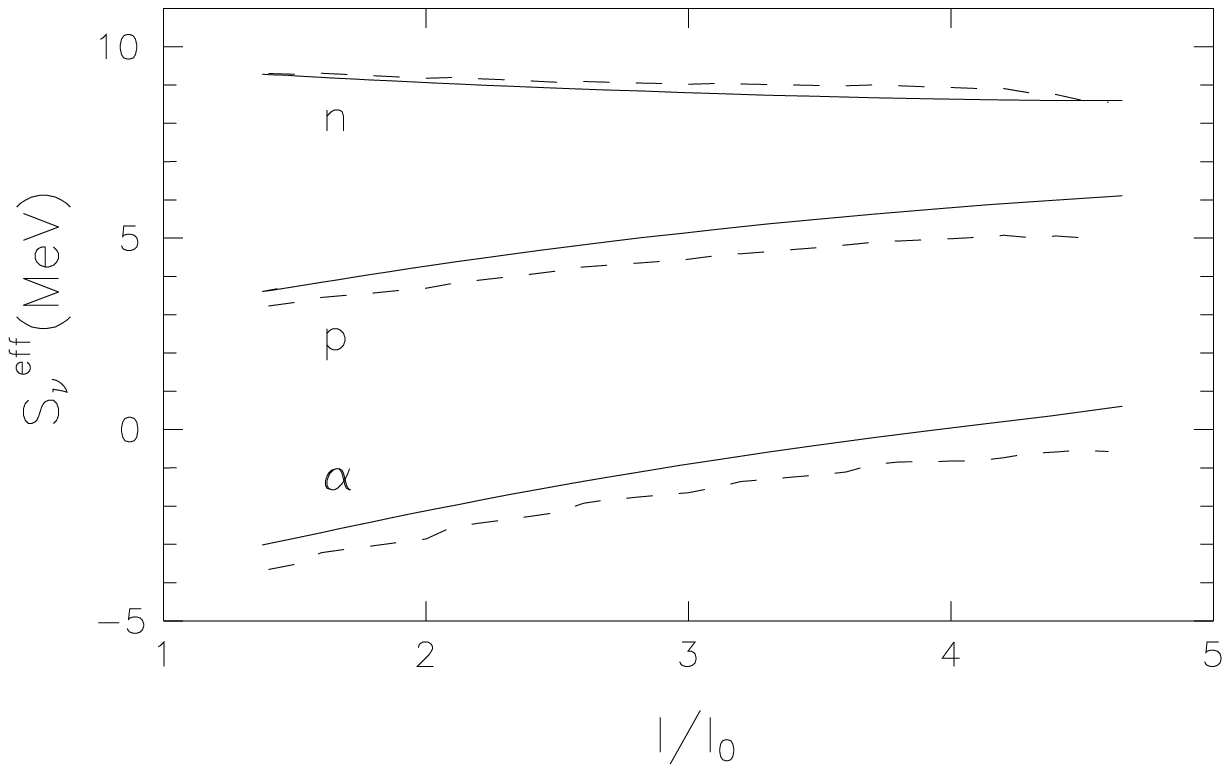}

\vspace*{1cm} Figure 7
\end{center}

\vspace*{1cm}

\begin{center}
\includegraphics{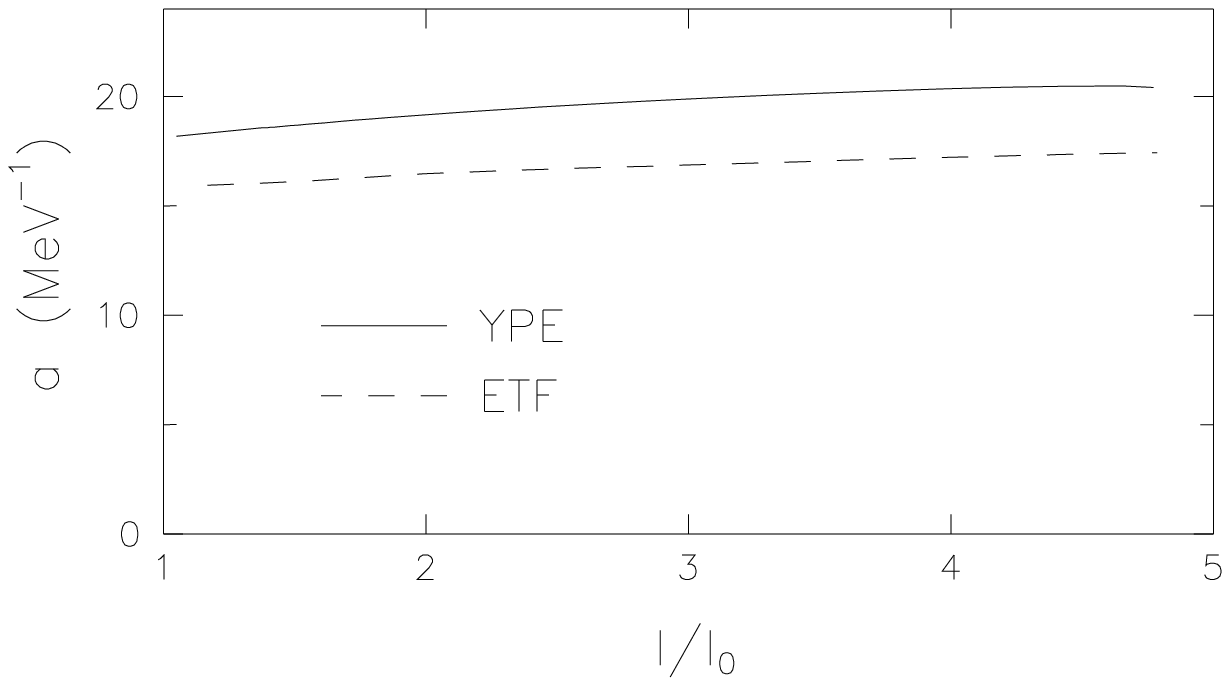}

\vspace*{1cm} Figure 8
\end{center}

\begin{center}
\includegraphics{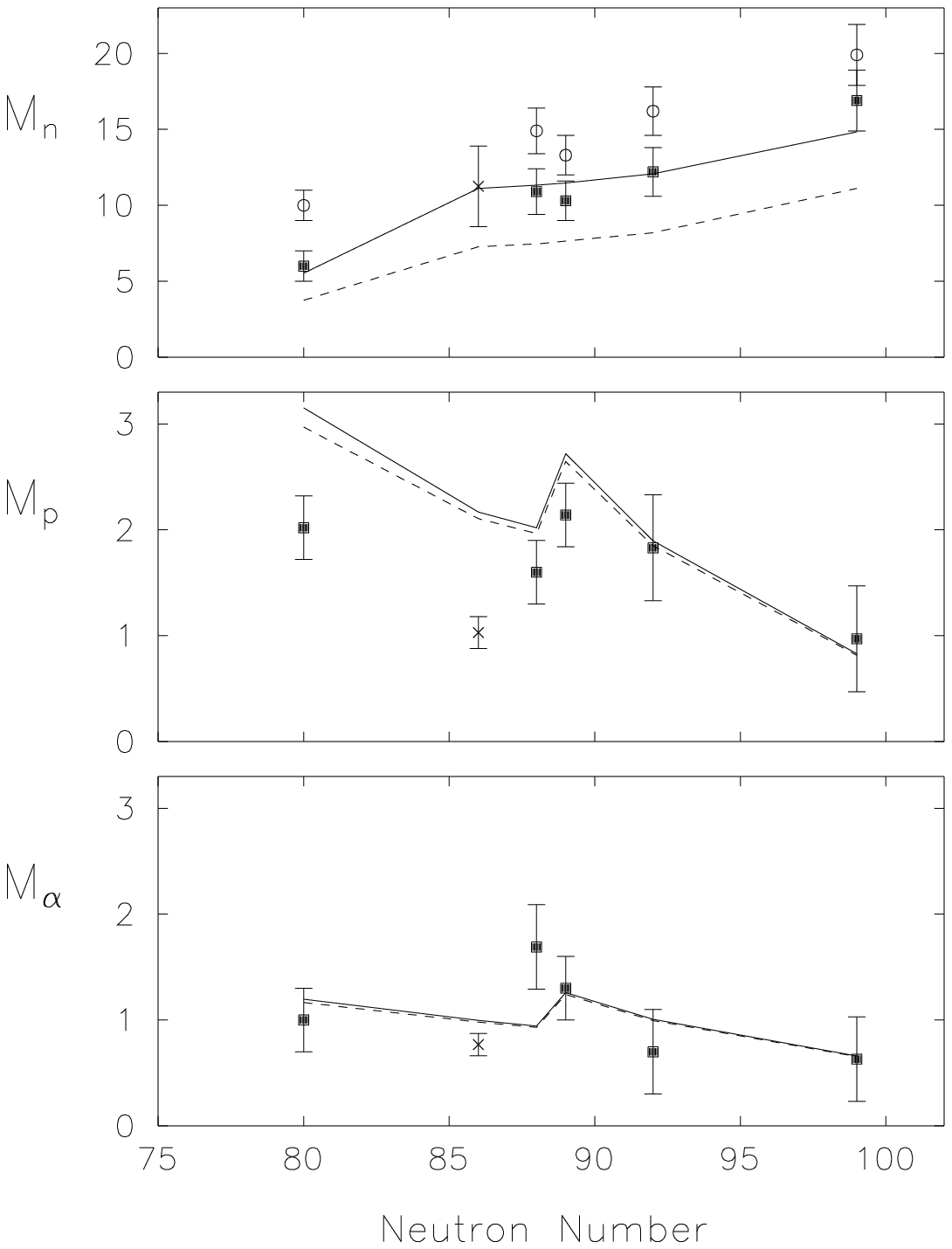}

\vspace*{2cm} Figure 9
\end{center}

\end{document}